\documentstyle[12pt]{article}

\begin{document}

\pagestyle{plain}
\setcounter{page}{1}
\setcounter{footnote}{00}
\renewcommand{\thefootnote}{\alph{footnote}}

\baselineskip=18pt 
\def\doublespaced{\baselineskip=\normalbaselineskip\multiply
    \baselineskip by 150\divide\baselineskip by 100}

\def\beq{\begin{equation}}
\def\enq{\end{equation}}
\def\su{$SU(2)_{\em l} \times SU(2)_h\times U(1)_Y$\,}
\def\uem{$U(1)_{\rm{em}}$\,}
\def\suu{$SU(2)\times U(1)_Y$\,}
\def\beq{\begin{equation}}
\def\enq{\end{equation}}
\def\ra{\rightarrow}
\def\D0{D\O~}
\def\ETslash{\not{\hbox{\kern-4pt $E_T$}}}
%
%
\begin{titlepage}
\baselineskip=0.3in
\begin{flushright}
PKU-TP-97-05\\
NUHEP-TH-96-7\\
MSUHEP-61122
\end{flushright}
\vspace{0.2in}
\begin{center}
{\Large Supersymmetric Electroweak Parity Nonconservation\\
      in Top Quark Pair Production at the Fermilab Tevatron }\\
\vspace{.2in}
Chong Sheng Li$^{(a)}$,  Robert J. Oakes$^{(b)}$, Jin Min Yang$^{(b,c)}$,
and C.--P. Yuan$^{(d)}$ \\
\vspace{.2in}
${(a)}$ Department of Physics, Peking University, China\\
${(b)}$ Department of Physics and Astronomy, Northwestern University,\\
     Evanston, Illinois 60208-3112, USA\\
${(c)}$ International Institute of Theoretical and Applied Physics,\\
        Iowa State University, Ames, IA 50011, USA\\
${(d)}$ Department of Physics and Astronomy, Michigan State University,\\
        East Lansing, Michigan 48824, USA\\
 

\end{center}
\vspace{.3in}

\begin{center}\begin{minipage}{5in}
\baselineskip=0.25in
\begin{center} ABSTRACT\end{center}

We evaluate the supersymmetric (SUSY) electroweak corrections to the effect 
of parity nonconservation in $ q \bar q \ra t \bar t$ production at the
Fermilab Tevatron predicted by the Minimal Supersymmetric Model (MSSM).
We find that the parity nonconserving asymmetry from the SUSY electroweak 
and SUSY Yukawa loop corrections predicted by the minimal supergravity 
(mSUGRA) model and the MSSM models with scenarios motivated by current data 
is about one percent. It will be challenging to observe such a small
asymmetry at the Tevatron with $10\,{\rm fb}^{-1}$ of luminosity.
It could however be observable if both the top- and bottom-squarks are 
light and $\tan \beta$ is smaller than 1, though theses parameters are 
not favored by mSUGRA. 
\end{minipage}\end{center}
\vspace{.5in}


\end{titlepage}
\newpage

\baselineskip=18pt 
\renewcommand{\thefootnote}{\arabic{footnote}}
\setcounter{footnote}{0}

\section{ Introduction }
\indent

Because of its large mass,
$m_t=175\pm 6$\,GeV \cite{topdisc},
the lifetime of the top quark [$\sim {(1.5\,{\rm GeV})}^{-1}$]
is shorter than the time [$\sim {(0.2\,{\rm GeV})}^{-1}$]
needed to flip its spin. Hence, it is possible to study the
polarization of this bare quark from its weak decay \cite{toppol}.
Using this unique property of the top quark,
it was suggested in Ref.~\cite{toppv} that the asymmetry 
(denoted as ${\cal A}$) in the 
production rates of left-handed and right-handed top quarks 
observed at the Tevatron
could be a good observable for probing physics that induces
parity nonconserving effects.

Because QCD is C (charge conjugation) and P (parity) invariant, the
single particle polarization
 of the top quark has to vanish 
for the QCD process $q \bar q , gg \ra t \bar{t}$,\footnote{
Here, we shall not consider the degree of 
(transverse) polarization of the top 
quark in the direction that is perpendicular to the
scattering plan of the $t \bar t$ pair production. This transverse
polarization can be generated through loop effects from QCD interaction 
\cite{toptloop,toppol}.
}
but can be nonvanishing if weak effects are
present in their production. The contribution of the Born level 
electroweak reaction 
$q\bar{q} \ra (\gamma,Z) \ra t\bar{t}$
to the total cross section for the
production of $t\bar{t}$ pairs is small (at the percent level for the
Tevatron), so any spin effects present in 
$q\bar{q}\ra(\gamma,Z)\ra t\bar{t}$
are diluted by the QCD
production of $t\bar{t}$ pairs.  Considering the larger rate for the
QCD production of $t\bar{t}$ pairs, similar spin effects that appear when
considering the degree of polarization due to the weak
corrections to $q\bar{q}\to g\to t\bar{t}$ at the loop level can be more
significant.
It was found in Refs.~\cite{toppv} and \cite{toppvtwo} 
that the effect from the Standard Model (SM) weak corrections
to this asymmetry is typically less than a fraction of percent.
It increases as the invariant mass ($M_{t \bar t}$)
of the $t \bar t$ pair increases
and reaches about 0.4\% for $M_{t \bar t}>500$\,GeV and 
about 1\% at the TeV region.

Because the SM contribution is small, 
this asymmetry may provide a good probe for new physics beyond the SM. 
One of the candidates for new physics consistent with current low
energy data (including $Z$-pole physics) is 
the Minimal Supersymmetric Standard Model (MSSM) \cite{hkmssm}.
Recently in the context of the MSSM the effect of
Yukawa contributions from
the Higgs sector to this asymmetry ${\cal A}$ 
in the $q \bar q \ra t \bar t$
process was evaluated and found to be of the same size as the SM 
effect for $\tan\beta \equiv v_2/v_1 < \sqrt{m_t/m_b}$
and enhanced by about a factor of 2  
for $\tan\beta > m_t/m_b$ when the mass of the charged Higgs boson 
$m_{H^+} < 300$\,GeV ~\cite{toppvtwo}.
However, in the MSSM, in addition to
 the Yukawa contributions from the Higgs 
sector, the genuine supersymmetric electroweak (SUSY EW) corrections from 
the spin-$1/2$ supersymmetric particles contributing in loops
should also be considered. 
This is because in the SUSY models the loop contribution
from spin-$0$ particles is often cancelled by that from spin-$1/2$ 
particles.
A complete study of the parity nonconserving effect in top 
quark pair production from SUSY models
should therefore include both the SUSY EW and SUSY Yukawa corrections.

In this work, we evaluate the 
genuine SUSY EW contribution of order $\alpha m_t^2/m_W^2$ 
to the parity nonconserving asymmetry ${\cal A}$ in the 
$t \bar t$ pair productions at the Fermilab Tevatron.
At the Tevatron, a ${\rm p}{\bar {\rm p}}$ collider with 
CM energy $\sqrt{s}=2$\,TeV, the $t \bar t$ pairs are 
produced predominately via the QCD process $q \bar q \ra t \bar t$;
 hence,  
we shall only consider the SUSY EW corrections 
for the $q \bar q \ra t \bar t$ process.  
With a 2\,${\rm fb}^{-1}$ luminosity (expected 
at the Tevatron Run II), there will be about 1000 fully reconstructed 
$t \bar t$ pairs \cite{tev2000}, which implies that 
it is possible to observe the parity nonconserving asymmetry of $\sim 3 \%$
in magnitude.
To probe physics that contributes to a smaller $|{\cal A}|$,
a higher luminosity is needed. For completeness,
we shall also give our results for an integrated luminosity of 
10\,${\rm fb}^{-1}$.

\section{SUSY electroweak corrections}
\indent

The genuine supersymmetric electroweak contribution
of order $\alpha m_t^2/m_W^2$ to the amplitude 
of $q \bar q \ra t \bar t$
is contained in the radiative corrections to the $t$-$\bar t$-$g$ vertex.
The  relevant Feynman diagrams were 
shown in Fig. 1 of Ref. \cite{csli1}, and the  
 Feynman rules can be found in Ref. \cite{hkmssm}.
In our calculation, we used dimensional regularization to regulate all the 
ultraviolet divergences in the virtual loop corrections and we adopted the 
on-mass-shell renormalization scheme \cite{renor}.
The renormalized $gt\bar t$ vertex is 
\begin{equation}
	 -i g_s \bar u(p)T^a [\Gamma_0^{\mu}+\Gamma_1^{\mu}] v(q),
\end{equation}
where $p,q$ are the momenta of outgoing $t$ quark and its antiparticle, 
$\Gamma_0^{\mu}=\gamma^{\mu}$ is the tree level vertex and
$\Gamma_1^{\mu}$ is the one-loop vertex function which can be expressed 
in terms of form factors \cite{toppol}
\begin{eqnarray}
\Gamma_1^{\mu}&=&\gamma^{\mu}[A(\hat s)-B(\hat s)\gamma_5]\nonumber\\
& & + (p-q)^{\mu}[C(\hat s)-D(\hat s)\gamma_5]\nonumber\\
& & + (p+q)^{\mu}[E(\hat s)-F(\hat s)\gamma_5],
\end{eqnarray}
where $\hat s=(p+q)^2$.
If CP is conserved, then $D=0$.\footnote{
In the SM, this is the case when ignoring the CP violating phase 
in the Cabibbo-Kobayashi-Maskawa quark mixing matrix.
In SUSY models, it is possible to have other CP violating sources,
such as the trilinear term ($A$-term) in the Higgs potential. 
However, for simplicity, we shall not include them in this study. 
} 
The conservation of the vector current for the 
renormalized $gt\bar t$ vertex demands that $E=0$ and 
$F({\hat s})=-2 m_t B({\hat s}) / {\hat s} $. 

The form factors from the supersymmetric electroweak corrections
can be written as
\begin{eqnarray}
A&=&\frac {g^2 m_t^2}{32\pi^2 m_W^2 \sin^2 \beta} 
(F_1+2m_t F_5),\nonumber\\
B&=&-\frac {g^2 m_t^2}{32\pi^2 m_W^2 \sin^2 \beta} F_2,\nonumber\\
C&=&-\frac {g^2 m_t^2}{32\pi^2 m_W^2 \sin^2 \beta} F_5,\nonumber\\
D&=&E=0,\nonumber\\
F&=&- \frac {2 m_t} {\hat s}  B({\hat s}) , 
\label{forms}
\end{eqnarray}
where $F_{1,2,5}$ are defined as in Ref.~\cite{csli2} by
\begin{equation}
F_i=F_i^c+F_i^n
\label{ficn} \, ,
\end{equation}
and $F_i^c$ and $F_i^n$ are the loop contributions 
 from diagrams involving charginos and neutralinos, 
respectively. Also \cite{csli2},
\begin{eqnarray}
F_1^c&=&\sum_{j=1,2}V_{j2}V_{j2}^* \left [ c_{24}+m_t^2(c_{11}+c_{21})
+(\frac{1}{2}B_1+m_t^2 B'_1)(m_t,\tilde M_j,m_{\tilde b})\right ],
\nonumber\\
F_2^c&=&\sum_{j=1,2}V_{j2}V_{j2}^* \left [ c_{24}
      +\frac{1}{2}B_1(m_t,\tilde M_j,m_{\tilde b}) \right ],\nonumber\\
F_5^c&=&-\frac{1}{2}m_t \sum_{j=1,2}V_{j2}V_{j2}^*
	\left ( c_{11}+c_{21} \right ),\nonumber\\
\end{eqnarray}
In the above results,
 the scalar functions $c_{ij}(-p_3,p_3+p_4,\tilde M_j, m_{\tilde b},
m_{\tilde b})$ and $B_1$ are the usual 3-point and 2-point
Feynman integrals, respectively \cite{loopint}.
The chargino masses $\tilde M_j$  and the mixing matrix elements 
$V_{ij}$ depend on the SUSY parameters $M_2$, $\mu$, and 
$\tan\beta$ \cite{hkmssm}.
Furthermore, $B'_{0,1}$ are defined by 
\begin{equation}
B'_{0,1}(m,m_1,m_2)=
\frac{\partial B_{0,1}(p,m_1,m_2)}{\partial p^2}
				\vert_{p^2=m^2}.
\end{equation}
In Eq.~(\ref{ficn}), $F_i^n$ can be written as 
\begin{eqnarray}
F_i^n&=&F_i^{\tilde t_1}+F_i^{\tilde t_2}+F_i^s ,
\label{fni}
\end{eqnarray}
where $F_1^s$ and $F_2^s$ are given by  
\begin{eqnarray}
F_1^s&=&\sum_{j=1}^4 \left \{ \frac{1}{2}N_{j4}N_{j4}^* \left [
B_1(m_t,\tilde M_{0j},m_{\tilde t_1})+
B_1(m_t,\tilde M_{0j},m_{\tilde t_2})
\right ] \right. \nonumber\\
& & +m_t^2 N_{j4}N_{j4}^* \left [
B'_1(m_t,\tilde M_{0j},m_{\tilde t_1})+
B'_1(m_t,\tilde M_{0j},m_{\tilde t_2})
\right ]  \nonumber\\
& & \left. +m_t \tilde M_{0j} N_{j4}N_{j4} \sin(2\theta_t) \left [
B'_0(m_t,\tilde M_{0j},m_{\tilde t_2})-
B'_0(m_t,\tilde M_{0j},m_{\tilde t_1})
\right ] \right \}, \nonumber\\
F_2^s&=&\sum_{j=1}^4 \frac{1}{2}N_{j4}N_{j4}^* \cos(2\theta_t)
\left [ B_1(m_t,\tilde M_{0j},m_{\tilde t_1})
 -B_1(m_t,\tilde M_{0j},m_{\tilde t_2}) \right ] \nonumber\\ 
F_5^s&=&0 \, ,
\end{eqnarray}
and $F_i^{\tilde t_1}$ are given by
\begin{eqnarray}
F_1^{\tilde t_1}&=&\sum_{j=1}^4 \left\{ N_{j4}N_{j4}^* \left [ c_{24}
    		 +m_t^2(c_{11}+c_{21})\right ]
    -\sin(2\theta_t) N_{j4}N_{j4} m_t \tilde M_{0j}(c_0+c_{11})\right \},
\nonumber\\
F_2^{\tilde t_1}&=&\sum_{j=1}^4 N_{j4}N_{j4}^* c_{24}\cos(2\theta_t),
\nonumber\\
F_5^{\tilde t_1}&=&\sum_{j=1}^4 \left [ 
-\frac{1}{2} m_t N_{j4}N_{j4}^* ( c_{11}+c_{21} )
+\frac{1}{2}\sin(2\theta_t) 
N_{j4}N_{j4} \tilde M_{0j}(c_0+c_{11})\right ],
\end{eqnarray}
where $c_{ij}(-p_3,p_3+p_4,\tilde M_{0j}, m_{\tilde t_1},m_{\tilde t_1})$
and $c_0$
are the 3-point Feynman integrals \cite{loopint}. 
The neutralino masses $\tilde M_{0j}$ 
and the mixing matrix elements $N_{ij}$ are obtained by diagonalizing 
the neutralino mass matrix $Y$ \cite{hkmssm} which depends on   
the SUSY  parameters $M_1$, $M_2$, $\mu$ and $\tan\beta$.
Here, $\mu$ is the coefficient of
the $H_1H_2$ mixing term in the superpotential and 
$M_2, M_1$ are the  $SU(2)$ and the $U(1)$
gaugino masses, respectively.
In general, the mass matrix of top-squarks takes the form \cite{stopm}
\begin{equation}
{\cal M}_{\tilde t}^2 = \left ( \begin{array}{cc}
m^2_{\tilde t_L} & m_t M_{LR}\\m_t M_{LR} & m^2_{\tilde t_R} 
\end{array}\right ) \, ,
\end{equation}
where $ M^2_{\tilde t_L}, M^2_{\tilde t_R}$ are the soft SUSY-breaking mass
terms for left- and right-handed top-squarks, and 
$M_{LR}=\mu\cot\beta+A_t$, with $A_t$ being the 
coefficient of the trilinear soft SUSY-breaking term
$\tilde t_L\tilde t_R H_2$. 
The mass eigenstates of the top-squarks ${\tilde t_{1,2}}$ are
related to ${\tilde t_{L,R}}$ by
${\tilde t_{1}}= \cos \theta_t {\tilde t_{L}} 
+ \sin \theta_t {\tilde t_{R}}$, and 
${\tilde t_{2}}= - \sin \theta_t {\tilde t_{L}} 
+ \cos \theta_t {\tilde t_{R}}$,
where  $\theta_t$ is the mixing angle of the top-squarks.

The form factors $F_i^{\tilde t_2}$ can be obtained
from  $F_i^{\tilde t_1}$ by a proper transformation, and
\begin{equation}
F_i^{\tilde t_2}=F_i^{\tilde t_1}\left\vert_{
 	\sin(2\theta_t)\rightarrow -\sin(2\theta_t),
	\cos(2\theta_t)\rightarrow -\cos(2\theta_t),
	m_{\tilde t_1}\rightarrow m_{\tilde t_2}} \right. \, .
\end{equation}
In the next section, we will discuss the formalism for calculating 
the asymmetry ${\cal A}$ which measures the degree of parity nonconservation
in $t \bar t$ pairs produced at the Tevatron.

\section{Asymmetry and parity nonconservation}
\indent

Although a  large part of this section is contained in Refs. \cite{toppv}
and \cite{toppvtwo}, for completeness, we will briefly summarize the 
relevant results.
Let us define the cross section for the subprocess $q\bar{q} \to t\bar{t}$
for definite helicity states of the $t\bar{t}$ pair to be
\begin{equation}
\hat{\sigma}_{\lambda_1,\lambda_2}
\equiv \hat{\sigma}( q\bar{q} \to t_{\lambda_1} \bar{t}_{\lambda_2} ),
\end{equation}
where $\lambda_{1,2}$ designates right-handed ($R$)
or left-handed ($L$) helicity.

At the tree level,
\begin{eqnarray}
\hat{\sigma}^{(0)}_{LL} & = & \hat{\sigma}^{(0)}_{RR}
 =   \frac{ 4 \pi \alpha_s^2 \beta}{27\hat{s}^2} (2m_t^2), \nonumber \\
\hat{\sigma}^{(0)}_{LR} & = & \hat{\sigma}^{(0)}_{RL}
 =   \frac{ 4 \pi \alpha_s^2 \beta}{27\hat{s}^2} (\hat{s}), \nonumber \\
\hat{\sigma}^{(0)}
 & = & \hat{\sigma}^{(0)}_{LL} +\hat{\sigma}^{(0)}_{RR}
      +\hat{\sigma}^{(0)}_{LR} +\hat{\sigma}^{(0)}_{RL}
   =  \frac{ 8 \pi \alpha_s^2 \beta}{27\hat{s} }( 1+2m_t/\hat{s} )
\label{cross} \, ,
\end{eqnarray}
after properly including the spin and color factors.

With one loop radiative corrections the cross section
for each helicity state of the $t\bar{t}$ pair is
$\hat{\sigma}_{\lambda_1,\lambda_2} =  
\hat{\sigma}^{(0)}_{\lambda_1,\lambda_2}
+ \delta \hat{\sigma}_{\lambda_1,\lambda_2}$,
where $\delta \hat{\sigma}_{\lambda_1,\lambda_2}$
is the contribution from weak corrections.
Defining a $K$-factor for each helicity state of the $t\bar{t}$ pair to be
\begin{eqnarray}
K_{\lambda_1,\lambda_2}
& \equiv & \frac{\hat{\sigma}_{\lambda_1,\lambda_2}}
                {\hat{\sigma}^{(0)}_{\lambda_1,\lambda_2}}
        =  1 +\frac{\delta \hat{\sigma}_{\lambda_1,\lambda_2}}
                   {\hat{\sigma}^{(0)}_{\lambda_1,\lambda_2}} \, ,
\end{eqnarray}
these $K$-factors are 
\begin{eqnarray}
K_{LL}& = &1 +2Re(A) -\beta^2 \hat{s} Re(C)/m_t +\beta \hat{s} Re(D)/m_t
           \nonumber , \\
K_{RR}& = &1 +2Re(A) -\beta^2 \hat{s} Re(C)/m_t -\beta \hat{s} Re(D)/m_t
           \nonumber , \\
K_{LR}& = &1 +2Re(A) +2\beta Re(B)
           \nonumber , \\
K_{RL}& = &1 +2Re(A) -2\beta Re(B) \, ,
\end{eqnarray}
with $\beta = \sqrt{ 1 -4m_t^2/\hat{s} }$.
If CP is conserved, then $Re(D)=0$ and $K_{LL}=K_{RR}$.
Because $t_R {\bar t}_L$ is the conjugate state of 
$t_L {\bar t}_R$ under parity transformation, any difference 
in $K_{RL}$ and $K_{LR}$ implies a parity nonconserving interaction.

The cross section for $p\bar{p} \to t\bar{t} +X$
is evaluated by calculating the convolution
of the constituent cross section $\hat{\sigma}(q\bar{q} \to t\bar{t})$
and parton distribution functions.
The effects of parity nonconservation can appear as an asymmetry
in the invariant mass ($M_{t {\bar t}}$) distributions as well as
in the integrated cross sections for $t_R\bar{t}_L$ and $t_L\bar{t}_R$
productions.
Let us define the differential asymmetry to be
\begin{eqnarray}
\delta {\cal A}(M_{t\bar{t}})
& \equiv &\frac{d\sigma_{RL}/dM_{t\bar{t}} -d\sigma_{LR}/dM_{t\bar{t}}}
          {d\sigma_{RL}/dM_{t\bar{t}} +
d\sigma_{LR}/dM_{t\bar{t}}} \nonumber \\
& = &\frac{ K_{RL}-K_{LR} }{ K_{RL}+K_{LR} } \nonumber \\
& = &\frac{-2\beta Re(B) }{ 1 +2Re(A) },
\end{eqnarray}
where the parton distribution functions cancel in this ratio.

Since the polarization of the top quark has to be deduced from the 
distributions of the decay products, such as $b$-jets and leptons,
it is necessary to have a fully reconstructed $t \bar t$ event
to determine the polarizations of both $t$ and $\bar t$.
To increase statistics we can sum over the helicities of the $\bar{t}$
and consider an integrated asymmetry in the
numbers of $t_R$ and $t_L$ produced.
In this case, the number of observed $t\bar{t}$ events
will be reduced by only
the branching ratio for $t \to W^+ b \to \ell^+\nu_\ell b$ , where
$\ell=e \, {\rm or} \, \mu$.
The integrated asymmetry, after integrating over a range of 
$M_{t \bar t}$, is then defined by
\begin{eqnarray}
{\cal A}
  & \equiv &\frac{ N_{R} -N_{L} }{ N_{R} +N_{L} }
     = \frac{ \sigma_{R} -\sigma_{L} }{ \sigma_{R} +
\sigma_{L} },
\end{eqnarray}
where $\sigma_{R} = \sigma_{RL} +\sigma_{RR}$,
$\sigma_{L} = \sigma_{LR} +\sigma_{LL}$,
and $N_R$ and $N_L$ are the numbers of right-handed and
left-handed top quarks which decay semileptonically.

\section{ Numerical results and conclusions }
\indent

 In our numerical calculations, to facilitate comparison
with the effect from the SUSY Yukawa corrections evaluated
in \cite{toppvtwo}, 
we take the same values as those in Ref. \cite{toppvtwo}
 for the SM parameters.
More precisely, we take 
$m_Z=91.187$\,GeV,  
$m_W=80.22$\,GeV, $m_b=4.8$\,GeV, $m_t=170$\,GeV, 
and $\sin^2\theta_W=0.2319$.
We also use the same parton distribution functions, CTEQ3L \cite{cteq3},
to evaluate the cross section with the choice of scale
$Q^2=\hat s=M^2_{t \bar t}$. 
As to the SUSY parameters we shall consider two classes
of models in order.

The currently most popular SUSY model is the minimum supergravity
(mSUGRA) model \cite{sugra} which only has four continuous and one discrete 
free parameters not present in the SM. 
At the unification scale $M_X$ these are
$m_0$ (common scalar mass), $m_{1/2}$ (common gaugino mass),
$A_0$ (common trilinear scalar coupling), $\tan \beta$ (the ratio of
vevs), and sign($\mu$).
A common value $m_{1/2}$ at the scale $M_X$ 
is motivated by the apparent unification 
of the measured gauge couplings within the MSSM at the 
scale $M_X \simeq 2 \times 10^{16}$\,GeV.
Through the renormalization group equations of the mSUGRA model,
the gaugino masses $M_1$ and $M_2$ are related by 
$M_1=\frac{5}{3}\frac{{g'}^2}{g^2} M_2 \simeq 0.5 M_2$
at the electroweak scale.
This model predicts radiative breaking of the electroweak gauge symmetry
due to the large top quark mass.
Consequently, it is possible to have large splitting in the masses of
left-top squarks and right-top squarks, while the masses of all the other 
(left- or right-) squarks are about the same \cite{kanesugra}.
Because of the heavy top quark this model requires values of 
$\tan \beta$ to be larger than $\sim 1$ so that the 
top Yukawa coupling will not become exponentially large below
the high energy scale $M_X$, especially since there is evidence that
the gauge couplings are perturbative up to that scale.
Similarly, $\tan \beta$ is required
to be smaller than $\sim 50$ where the bottom and $\tau$ 
Yukawa couplings are likewise close to their perturbative 
limits \cite{kanesugra}.

Another type of SUSY model uses the full ($> 100$ parameters)
parameter space freedom of the MSSM and fits the data, assuming one
has a supersymmetry signal. This approach has been used in
 studying the CDF $e^-e^+ \gamma \gamma + \ETslash $ event 
\cite{kanelight}.
It was found that to also explain all the low energy data
(including $\alpha_s$, $R_b$ and the branching ratio of
$b \rightarrow s \gamma$, etc.),
the lightest mass eigenstate ($\tilde t_1$) of top squarks 
is likely to be the right-stop ($\tilde t_R$) with a mass of the 
order of $m_W$; 
the lightest neutralino (${\tilde \chi}^0_1$) is 
Higgsino-like and the second 
 lightest neutralino (${\tilde \chi}^0_2$) is gaugino-like; 
$\tan \beta$ is of the order one; $M_2$ and $M_1$ are 
of the same order of magnitude 
as $m_Z$ (the mass of $Z$-boson); the sign($\mu$) is negative; and 
$|\mu| \sim M_Z$ \cite{kanelight}.
We shall refer to this class of models as MSSM models with 
scenarios motivated by current data.

The relevant SUSY parameters for our study are 
$\mu$, $\tan \beta$, $M_1$, $M_2$ (which determine couplings,  
masses and  mixings 
of neutralinos and charginos), and the masses of the
top- and bottom-squarks.
When considering SUSY EW corrections, charginos and neutralinos will 
contribute only in loops and their contributions 
are generally  small unless  they are light.
Likewise, only a light top-squark (or bottom-squark) can give large
contributions to loop calculations.
Since we are interested in  contributions of the
order $m_t^2/v^2$, where $v$ is the vacuum expectation value,
only Higgsino components of the charginos and the neutralinos 
are relevant for our calculation.
Because in the MSSM the top quark obtains its mass by interacting 
 with the second Higgs doublet through Yukawa coupling,
the coupling of Higgsino-top-stop is proportional to 
${m_t / v_2} = {m_t / v \sin \beta }$. 
Hence, we expect large parity  nonconserving asymmetry occurs
for small values of $\tan \beta$.

After sampling a range of values of SUSY parameters in  the
region that might give large contributions to 
the asymmetry ${\cal A}$, and which are also 
consistent with either of the above two classes of models,
we found that the asymmetry ${\cal A}$ is generally small,
less than a couple of percent in magnitude, though it can be 
either positive or negative depending on the values of 
the SUSY parameters. 
To illustrate the effects of SUSY EW corrections to the
parity nonconserving observable ${\cal A}$,  in Tables 1 and 2 we give a 
few representative sets of SUSY parameters and the corresponding
asymmetries ${\cal A}_{EW}$ (due to SUSY EW contributions)
 and/or ${\cal A}_{Y}$ (due to SUSY Yukawa contributions).  
The relevant parameters used for calculating the results
in Tables 1 and 2 are:
$M_{t {\bar t}} > 500$\,GeV;\footnote{
This is motivated by the conclusion 
in Refs. \cite{toppv} and \cite{toppvtwo} that $|{\cal A}|$ increases
as $M_{t \bar t}$ increases.
}
$m_{\tilde t_1}=50$\,GeV or $90$\,GeV;
$m_{\tilde t_2}=m_{\tilde b_1}=m_{\tilde b_2}=500$\,GeV;
$m_{H^+}=300$\,GeV, and $\mu=-60$\,GeV or $-90$\,GeV.
In addition, we choose the SUSY parameter 
$A_t=-\mu \cot \beta$ (i.e. $M_{LR}=0$) so that the 
mass eigenstates of the stops coincide
with $\tilde t_R$ and $\tilde t_L$ up to a phase.
Furthermore, we consider the case that $m_{\tilde t_R}$ is smaller 
than $m_{\tilde t_L}$ , which corresponds to $\theta_t=\pi/2$. 
Accordingly, the mass eigenstates of the stops 
${\tilde t_1}$ and ${\tilde t_2}$
are just ${\tilde t_R}$ (a right-handed stop) and $-{\tilde t_L}$
(a left-handed stop), respectively. 
Using these values we calculated the results for a 2\,TeV 
Tevatron with an integrated luminosity of 10\,${\rm fb}^{-1}$.
(The effect of parity nonconservation in the $t \bar t$  events
is too small to be observable with only 2\,${\rm fb}^{-1}$ of
luminosity.)

In Table 1, we show the parity nonconserving asymmetry
${\cal A}$  from SUSY EW contributions.
The representative results shown in this Table 
span the SUSY parameter space:
$1 < \tan \beta < 3 $, 
$ 50\,{\rm GeV} \, < M_1 < 100\,{\rm GeV}$,
$ 50\,{\rm GeV} \, < M_2 < 100\,{\rm GeV}$,
with $\mu=-60$ and $m_{\tilde t_1}=50$\,GeV or $90$\,GeV.
It is clear that the values of  $|{\cal A}_{EW}|$ 
can be of the same size for either $M_1 > M_2$,
$M_1 \sim M_2$ or $M_1 < M_2$, though the
asymmetry in the first case is slightly larger.
We have checked that for these sets of parameters
the mass of the lightest neutralino is about 40\,GeV 
(for $\tan \beta \sim 3$) to 60\,GeV (for $\tan \beta \sim 1$),
and the second lightest neutralino has a mass of about 60\,GeV to 80\,GeV;
hence, these models are consistent with the MSSM models 
with scenarios motivated by current data \cite{kanelight}.
The fact that the asymmetry is larger for smaller $\tan \beta$ 
can be understood as follows:  
First, because the Higgsino-top-stop coupling is proportional to  
$1/\sin \beta$, it becomes larger for smaller $\tan \beta$. 
Second,
for small $\tan \beta$ ($\sim 1$), ${\tilde \chi}^0_1$ is 
Higgsino-like and ${\tilde \chi}^0_2$ is gaugino-like 
for $M_1 \ge M_2$.\footnote{
Higgsino-like does not imply a hundred percent Higgsino component,
etc.
}
(For $M_1 < M_2$, the properties of
${\tilde \chi}^0_1$ and ${\tilde \chi}^0_2$ are reversed.)
Hence, we anticipate a larger asymmetry for a smaller $\tan \beta$
and  $M_1 \ge M_2$.
Furthermore, as anticipated, 
 a lighter stop gives a larger 
parity nonconserving asymmetry through loop corrections.
After choosing the sets of SUSY 
parameters that give the largest parity nonconserving effect, 
we found that the asymmetry ${\cal A}_{EW}$ is 
positive and of the order of a percent.
To observe such a small effect, one needs at 
least $10^{4}$ fully reconstructed $t \bar t$ pairs in the 
lepton plus jets decay channel, which is at the limit of what 
can be expected with 10\,${\rm fb}^{-1}$ of luminosity at the Tevatron. 

In the general mSUGRA model $\tan \beta$ can be much larger than 1.
For larger $\tan \beta$ it was shown that the parity nonconserving 
asymmetry from the SUSY Yukawa contributions can be
enhanced \cite{toppvtwo}.
As indicated in Table 2, for the sets of SUSY 
parameters that give large parity nonconservation, 
the asymmetry ${\cal A}_{EW}$ have the same sign 
as  ${\cal A}_{Y}$.
Also, ${\cal A}_{EW}$ is larger than ${\cal A}_{Y}$ 
for $\tan \beta \sim 1$,
smaller for $\tan \beta > m_t/m_b $, and 
about the same for $\tan \beta \sim 3$.
The sum of the asymmetries ${\cal A}_{EW}$ and ${\cal A}_{Y}$
is about one percent, and not sensitive to the 
detailed parameters of the SUSY models.

Recall that the SM prediction of the
asymmetry ${\cal A}$, at
the order $m_t^2/v^2$, is about 0.36\% 
for a 100\,GeV Higgs boson
\cite{toppv,toppvtwo}.
Since none of the asymmetries shown in Tables 1 and 2 is very large, 
being typically larger than a percent,
we conclude that to observe the parity 
nonconserving asymmetry ${\cal A}$ induced from the SUSY EW 
and SUSY Yukawa loop corrections predicted by the two
classes of models discussed above will be challenging  
at the Tevatron
with  $10\,{\rm fb}^{-1}$ of luminosity. 
However, because of the strong enhancement factor 
$1/\sin^2\beta$ coming from the square of the  Higgsino-top-stop coupling 
[cf. Eq. (\ref{forms})], the parity nonconserving asymmetry
$|{\cal A}|$ can easily be much larger 
if $\tan \beta$ is significantly less than unity. 
Although this is not favored by either of 
the two classes of models considered above, 
it is in principle allowed in the MSSM. 
Another possibility that gives a large parity nonconserving 
asymmetry in the MSSM is to have not only a light stop but also a light
sbottom; however, this is again not favored by either of the two classes of
models discussed above. 
Finally, we show in Table~3 the predicted asymmetries for the SUSY models
with $\theta_t=0$ in which the left-handed top-squark
is lighter than the right-handed one.
In this case, the asymmetry ${\cal A}_{EW}$ is negative while 
${\cal A}_{Y}$ is positive. Also, $|{\cal A}_{EW}|$ is larger 
than ${\cal A}_{Y}$ for $\tan \beta \sim 1$,
smaller for $\tan \beta > m_t/m_b $, and about the same for 
$\tan \beta \sim 3$. Hence, the contributions of ${\cal A}_{EW}$ 
and ${\cal A}_{Y}$  mostly cancel for $\tan \beta$ near 3.

\vspace{.5cm}

C.P.Y would like to thank G.L. Kane, C. Schmidt, and T. Tait 
for helpful discussions.
This work was supported in part by the U.S. Department of Energy, Division
of High Energy Physics, under Grant No. DE-FG02-91-ER4086, and
by the NSF Grant PHY-9507683.

\newpage



\begin{table}
\caption{ }
Parity nonconserving  asymmetries
in $p \bar p \ra t \bar t +X$ for $M_{t {\bar t}} > 500$\,GeV.
The CM energy of the collider was assumed to be 2\,TeV and its 
integrated luminosity  10\,${\rm fb}^{-1}$.
The other relevant SUSY parameters are  
$m_{\tilde t_2}=m_{\tilde b_1}=m_{\tilde b_2}=500$\,GeV,
$A_t=-\mu \cot \beta$, $\theta_t=\pi/2$, and $\mu=-60$\,GeV.

\vspace{0.1in}

\begin{center}
\begin{tabular}{|c|cc|c|c|}
\hline
$\tan \beta$ & $M_1$ & $M_2$ & 
${\cal A}_{EW}$ for $m_{\tilde t_1}=50$\,GeV &
${\cal A}_{EW}$ for $m_{\tilde t_1}=90$\,GeV \\
     & (GeV) & (GeV) & (\%)  & (\%)  \\
\hline
1.0  &  50  &  100  &  1.25    & 0.92  \\
     &  75  &   75  &  1.32    & 0.91  \\
     & 100  &   50  &  1.31    & 0.92  \\
\hline
1.5  &  50  &  100  &  0.89    & 0.61  \\
     &  75  &   75  &  0.95    & 0.59  \\
     & 100  &   50  &  0.96    & 0.63  \\
\hline
2.0  &  50  &  100  &  0.76    & 0.50  \\
     &  75  &   75  &  0.80    & 0.48  \\
     & 100  &   50  &  0.82    & 0.49  \\
\hline
3.0  &  50  &  100  &  0.66    & 0.42  \\
     &  75  &   75  &  0.69    & 0.41  \\
     & 100  &   50  &  0.71    & 0.41  \\
\hline
\end{tabular}
\end{center}
\end{table}


\begin{table}
\caption{ }
Parity nonconserving  asymmetries
in $p \bar p \ra t \bar t +X$ for $M_{t {\bar t}} > 500$\,GeV.
The CM energy of the collider was assumed to be 2\,TeV and its 
integrated luminosity  10\,${\rm fb}^{-1}$.
The other relevant SUSY parameters are 
$m_{\tilde t_1}=90$\,GeV, 
$m_{\tilde t_2}=m_{\tilde b_1}=m_{\tilde b_2}=500$\,GeV,
$m_{H^+}=300$\,GeV, $A_t=-\mu \cot \beta$,
$\theta_t=\pi/2$, and $\mu=-90$\,GeV.

\vspace{0.1in}

\begin{center}
\begin{tabular}{|c|cc|cc|c|}
\hline
$\tan \beta$ & $M_1$ & $M_2$ & ${\cal A}_{EW}$ & ${\cal A}_{Y}$ &
   ${\cal A}$ \\
     & (GeV) & (GeV) & (\%) & (\%) & (\%)    \\
\hline
1    &  75  &  150  &  0.74  &  0.16  & 0.90  \\
     & 100  &  100  &  0.72  &  0.16  & 0.88  \\
     & 150  &   75  &  0.73  &  0.16  & 0.89  \\
\hline
3    &  75  &  150  &  0.37  &  0.33 &  0.70  \\
     & 100  &  100  &  0.35  &  0.33 &  0.68  \\
     & 150  &   75  &  0.34  &  0.33 &  0.67  \\
\hline
10   &  75  &  150  &  0.32  &   0.37  & 0.69 \\
     & 100  &  100  &  0.30  &   0.37  & 0.67 \\
     & 150  &   75  &  0.30  &   0.37  & 0.67 \\
\hline
35   &  75  &  150  &   0.32 &  0.54   & 0.86 \\
     & 100  &  100  &   0.30 &  0.54   & 0.84 \\
     & 150  &   75  &   0.29 &  0.54   & 0.83 \\
\hline
\end{tabular}
\end{center}
\end{table}


\begin{table}
\caption{ }
Parity nonconserving  asymmetries
in $p \bar p \ra t \bar t +X$ for $M_{t {\bar t}} > 500$\,GeV.
The CM energy of the collider was assumed to be 2\,TeV and its 
integrated luminosity  10\,${\rm fb}^{-1}$.
The other relevant SUSY parameters are 
$m_{\tilde t_1}=90$\,GeV, 
$m_{\tilde t_2}=m_{\tilde b_1}=m_{\tilde b_2}=500$\,GeV,
$m_{H^+}=300$\,GeV, $A_t=-\mu \cot \beta$,
$\theta_t=0$, and $\mu=-90$\,GeV.

\vspace{0.1in}

\begin{center}
\begin{tabular}{|c|cc|cc|c|}
\hline
$\tan \beta$ & $M_1$ & $M_2$ & ${\cal A}_{EW}$ & ${\cal A}_{Y}$ &
   ${\cal A}$ \\
     & (GeV) & (GeV) & (\%) & (\%) & (\%)  \\
\hline
1  &  75  &  150  &  -0.58  &  0.16  & -0.42 \\
     & 100  &  100  &  -0.58 &   0.16 &  -0.42 \\
     & 150  &   75  & -0.57  &  0.16  & -0.41    \\
\hline
3  &  75  &  150  &  -0.28 &  0.33 &  0.05   \\
     & 100  &  100  & -0.26 &  0.33 &  0.07   \\
     & 150  &   75  &  -0.25 &  0.33 &  0.08   \\
\hline
10  &  75  &  150  & -0.25 &   0.37  & 0.12   \\
     & 100  &  100  &  -0.22  &  0.37 &  0.15   \\
     & 150  &   75  & -0.21  &  0.37  & 0.16   \\
\hline
35  &  75  &  150  &   -0.24  &  0.54  & 0.30    \\
     & 100  &  100  &  -0.22 &  0.54  & 0.32    \\
     & 150  &   75  &   -0.21  &  0.54  & 0.33   \\
\hline
\end{tabular}
\end{center}
\end{table}

\end{document}